\documentclass{article}

\usepackage[english]{babel}

\usepackage[letterpaper,top=2cm,bottom=2cm,left=3cm,right=3cm,marginparwidth=1.75cm]{geometry}

\newcommand\tradeoff{participation-allocation tradeoff}
\newcommand\articletitle{Heterogeneous participation and allocation skews: when is choice ``worth it''?}

\usepackage{graphicx,subcaption}
\usepackage{url}

\usepackage{xcolor}

\usepackage{natbib}
\usepackage{cleveref}
\let\cite\citep

\title{\articletitle}
\author{Nikhil Garg}

\begin{document}

\maketitle

\begin{abstract} 
A core ethos of the Economics and Computation (EconCS) community is that people have complex private preferences and information of which the central planner is unaware, but which an appropriately designed mechanism can uncover to improve collective decisionmaking. This ethos underlies the community's largest deployed success stories, from stable matching systems to participatory budgeting. I ask: is this choice and information aggregation ``worth it''? In particular, I discuss how such systems induce \textit{heterogeneous participation}: those already relatively advantaged are, empirically, more able to pay time costs and navigate administrative burdens imposed by the mechanisms. I draw on three case studies, including my own work -- complex democratic mechanisms, resident crowdsourcing, and school matching. I end with lessons for practice and research, challenging the community to help reduce participation heterogeneity and design and deploy mechanisms that meet a ``best of both worlds'' north star: \textit{use preferences and information from those who choose to participate, but provide a ``sufficient'' quality of service to those who do not.}

\end{abstract}

% \begin{bottomstuff} 
% Author's address: \texttt{ngarg@cornell.edu}
% \end{bottomstuff}
            
% \maketitle

\section{Introduction}

A deserved point of pride for the Economics and Computation (EconCS) community is the integration into everyday life the systems we have long studied, an integration often done in collaboration with researchers. In New York City, I recently voted in a participatory budgeting election and used ranked choice voting for a mayoral primary election; my neighbors submit preferences to stable matching processes that assign their children to 3-k (for three year olds), pre-k, kindergarten, middle school, and high school; and the city has embraced crowdsourcing: whenever we encounter problems as mundane as potholes or as serious as suspected lead in our water, we can submit a 311 report or request a testing kit. 
Each of these systems represents a triumph of an underlying community ethos: that the people have complex preferences and information of which the government is unaware, but which an appropriately designed mechanism can uncover to improve collective decisionmaking.

This article's purpose is to raise a simple, perhaps surprising, question: \textit{is this choice and information aggregation ``worth it''}? Just as democratic decisionmaking generally privileges those who (can) vote, these systems skew public resource allocation and decisionmaking in favor of those who (can) participate. And, as I will describe, substantial empirical evidence has established that participation in these mechanisms correlates with existing axes of privilege. Thus, we must ask whether the gain in information aggregation is worth the cost---or have we, in the guise of preference optimization, deployed ways to allocate scarce public resources to those best positioned to take advantage? As I will argue, this question is central to the legitimacy -- and perceived legitimacy -- of our systems. 

My thesis is analogous to, and motivated by, those recently advanced in policymaking, public interest technology,  and behavioral economics. In their seminal book, ``Administrative Burdens: Policymaking by Other Means,'' \citet{herd2019administrative} argue that the information  requirements  to access rights such as voting and Medicare -- often imposed in the name of safety, fraud detection, and choice -- in practice cause people to not receive what they are entitled to. In their respective books, \citet{schank2021power} and \citet{pahlka2023recoding} argue that poor technology design -- something as innocuous as long forms -- contributes to this loss, even when well-intentioned. In ``Scarcity: Why Having Too Little Means So Much,'' \citet{mullainathan2013scarcity} explain how poverty begets poverty, because it inhibits long-term planning in favor of urgent needs. All then argue that system designers must {design} with this phenomenon -- the time cost of participation -- in mind. Analogously, I argue that effective outcomes in the face of heterogeneous participation must be a primary design goal for our field, if we want our information aggregation mechanisms to be ``worth it.'' In other words, we should either deem ``equal''  participation as a necessary precondition to using choice to allocate scarce resources, or ensure that our mechanisms are robust despite heterogeneity.\footnote{I use the words, ``equal,'' ``heterogeneous,'' and ``representative'' informally. What exactly constitutes equal, or equal enough, depends on context and may be subjective. See \citet{chasalow2021representativeness} for a history and analysis of ``representativeness'' as a ``foundational yet slippery concept.''}

In this article, I first detail three case studies central to our community: complex voting mechanisms, resident crowdsourcing, and school matching. In each, I overview the promise and on-the-ground realities of how these systems affect collective decisionmaking. I highlight recent research, including my own, that has sought to understand and close the gap caused by heterogeneous participation. I then summarize shared patterns from the three case studies, including potential solutions and design principles. Finally, I overview practical and research directions on the use of choice to allocate scarce public resources. I challenge us to meet a ``best of both worlds'' north star: \textit{use preferences and information from those who participate, but provide a ``sufficient'' quality of service to those who do not.} Simultaneously, we should help develop approaches to support balanced participation. 

\section{Case studies}

\subsection{Complex democratic mechanisms}

``Equal'' voting rights and participation is central to democracy. Of course, equal participation is difficult to achieve; in the United States, eligible voters who are young, lower-income, racial and ethnic minorities, or have less formal education are less likely to vote \cite{hartig2023republican}. These patterns are also present in two democratic innovations advanced in the community: participatory budgeting and deliberative democracy (``citizen assemblies'').

In participatory budgeting, voters select which community projects to fund, from libraries in schools, to gym renovations, to park beautification. The Stanford Participatory Budgeting Platform has helped run over 150 elections, each of which may allocate millions of dollars \cite{gelauff2024rank}. New York City, Cambridge, Paris, Porte Alegro, Budapest, Helsinki, and many other cities globally all run participatory budgeting elections. At their best, these elections promise to increase civic engagement and ensure that project funding decisions are made by the people, instead of elected representatives or administrators. In deliberative democracy mechanisms, people (``panelists'') are selected to deliberate, potentially over several days, over a prescribed set of issues; they are polled before and after regarding their beliefs and sometimes are tasked to make recommendations; at their best, such processes gather a diverse set of people to make decisions in accordance with what a ``public sample would think if it had better conditions and information with which to explore and define the issues'' \cite{fishkin1991democracy}. Such processes have been used in over 25 countries, including to make constitutional amendments in Mongolia \cite{lee2024deliberative}, and EconCS researchers are involved in both building online deliberation platforms and in selecting panelists \cite{fishkin2019deliberative,flanigan2021fair}. 

Given time costs and the use of unfamiliar methods, ensuring representative participation in these processes is a continuous, challenging task, on which researchers have rightfully focused. Participatory budgeting is often conducted online and open to all residents, including children and non-citizens; however, turnout rates are sometimes low, including at or below 5\% of eligible voters, and there may be unequal participation rates by race, ethnicity, education, immigration status, and home ownership \cite{zepic2017participatory,stewart2014participatory,hayduk2017immigrant}. Others report that participatory budgeting increases civic engagement by otherwise disadvantaged groups \cite{johnson2023testing}, and there are mixed findings on its distributive effects \cite{shybalkina2019does,stewart2014participatory}.

 Motivated by unequal participation in participatory budgeting, \citet{gelauff2020advertising} and \citet{shen2021robust} study targeted advertising for demographically balanced participation. \citet{gelauff2022opinion} advocate for the design of ``civic feedback processes that are robust against disparities
in the representation of demographic and opinion minorities,'' including reweighting techniques ``for more equitable voice among demographic minorities
which were underrepresented in the process;'' such reweighting could especially be appropriate for processes which are consultative for policymakers as opposed to binding. 

Analogously, motivated by unequal volunteer and dropout rates in deliberative democracy, an important line of work shows how balanced panels can be selected (``sortition''). Both individual fairness (volunteers should have sufficiently high selection probabilities, even if from overrepresented groups) and overall representative balance (on both observed and unobserved covariates) are important \cite{benade2019no,flanigan2020neutralizing,flanigan2021fair,NEURIPS2021_d7b431b1,ebadian2022sortition,flanigan2023mini,baharav2024fair,10.1609/aaai.v38i9.28827,10.5555/3692070.3692290,10.5555/3709347.3743583,assos2025alternates}. Their algorithms have been deployed at scale to support panel selection \cite{flanigan2021fair}.

What lessons does this literature provide? (1) Representative participation is seen as a central design goal by researchers and practitioners, including in the EconCS community---it is well accepted that an unrepresentative process is not legitimate, though there is empirical equivocation on real-world participation disparities and its effect on resource allocation. (2) Representative participation (and overall rates) is nevertheless an ongoing challenge. 
In NYC participatory budgeting, fewer than 100,000 people vote city-wide,\footnote{The exact number of people eligible is unclear, as it depends on the current number of residents over 11 years old in the city council districts that participated. There are almost 8 million residents over 10 years old in NYC, implying a less than 2\% turnout rate.}  far less than even other local elections such as for city council---and some have questioned its use for decisionmaking  \cite{golliher2025participatory}. It thus remains unclear how to mitigate the effects of heterogeneous participation. %

\begin{figure}[tb]
	\centering
	\begin{subfigure}[c]{0.45\textwidth}
		\centering
		\includegraphics[width=\textwidth]{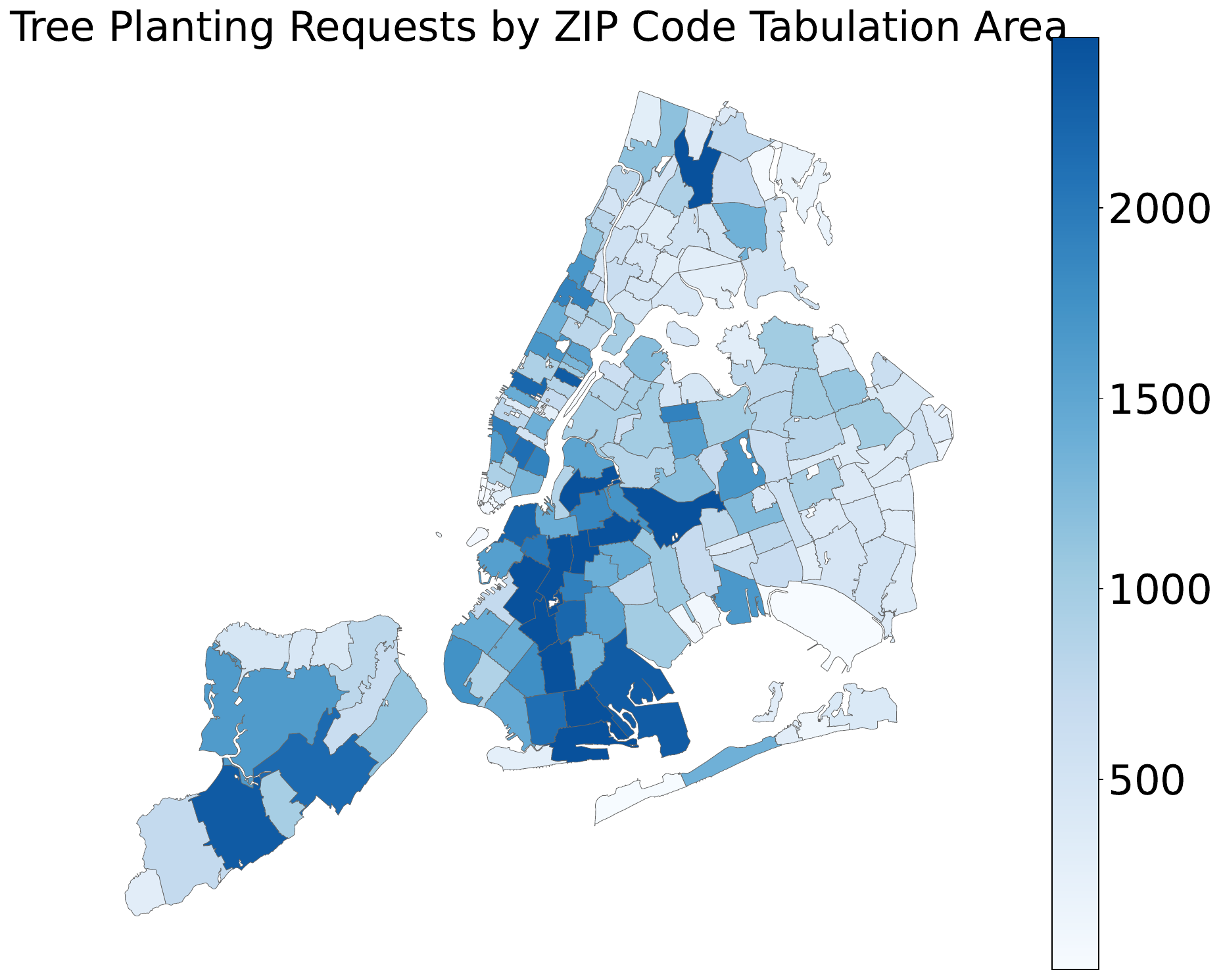}
		\label{fig:map}
	\end{subfigure}%
	\hfill
	\begin{subfigure}[c]{0.45\textwidth}
		\centering
			\begin{tabular}{p{2.5cm}rr}
			\hline
			\textbf{Variable} & \textbf{Coef} & \textbf{p-value} \\
			\hline
			Intercept & $-13720$ & $<0.001$ \\
			Log Population & $967$ & $<0.001$ \\
			Log Median Income & $472$ & $0.021$ \\
			Heat Vulnerability Index & $-149$ & $0.019$ \\
			\hline
		\end{tabular}
		\label{tab:regression}
	\end{subfigure}
	\caption{In NYC, the number of tree planting requests by ZIP Code Tabulation Area in 2015-2024. Controlling for population, requests correlate positively with neighborhood income and \textit{negatively} with a heat vulnerability index, a proxy for the need for shade. NYC no longer takes requests to plant trees, and instead will develop a schedule that prioritizes the most heat vulnerable areas.}
	\label{fig:treeplanting}
\end{figure}

\subsection{Resident crowdsourcing}
Another type of system deployed at scale to influence public resource allocation is \textit{resident crowdsourcing}: people make service requests, such as through ``311 systems'' in the United States. NYC receives over 3 million requests a year---for incidents ranging from fallen trees on powerlines, to potholes, flooding, rodents, and to request new tree planting---and similar systems are in place in hundreds of cities globally. This is an important avenue for the government to learn about problems -- supplementing and informing less frequent active inspections -- and there is a large government bureaucracy to respond to requests. 

However, substantial research, including my own, has established that participation is heterogeneous, even \textit{conditional on ground truth conditions}. For example, in \citet{liu2024quantifying}, we show how to use \textit{duplicate} reports about the same incident to estimate reporting delays; in \citet{agostini2024bayesian}, we use \textit{spatial correlation} to probabilistically identify unreported incidents; in \citet{balachandar2024using}, we combine regularly scheduled government inspections with crowdsourced reports; and in \citet{franchi2025bayesian}, we identify true flooding prevalence using a vision-language model on dashcams street imagery. In all cases, despite the diverse identification strategies, we find that (a) crowdsourced reporting data can be informative about ground truth conditions, e.g., that more hazardous conditions are reported at far higher rates \cite{liu2024quantifying}; but also (b) reporting is correlated with socioeconomic characteristics, also substantially: e.g., in \citet{liu2024quantifying}, we find that higher income, population density, voter participation, fraction of people with college degrees, and fraction of the population that is white all correlate with higher reporting rates. These patterns induce heterogeneous delays in incidents being addressed, potentially leading to inequitable government service.

What should we do given this heterogeneous participation? I do \textit{not} believe that these results imply we should not crowdsource this information; rather, we should ensure efficient resource allocation despite it, as in the work to balance citizen assemblies. For example, in a followup project, we seek to optimize inspection resources to efficiently and equitably set service level agreements \cite{liu2024redesigning}; it would be conceptually simple to account for heterogeneous reporting delays.

I believe that the design of such modifications is urgent, before practitioners decide that information aggregation is not worth the resulting allocation skews. NYC no longer allows the public to request new tree planting locations; instead, the Department of Parks and Recreation will plant ``street trees on a cyclical basis and prioritiz[e] the most heat-vulnerable neighborhoods first'' \cite{NYC311TreePlanting}. A simple analysis using public data \cite{NYCHeatVulnerabilityIndex,NYC311Data} helps explains why: as shown in \cref{fig:treeplanting}, planting requests historically correlated positively with median neighborhood income and \textit{negatively} with heat vulnerability, one  measure of ``need.'' Optimizing solely for stated resident preferences would lead to inefficient allocation, when the government has some expertise. Such pullback may also occur in other settings, if the mechanisms are not viewed as legitimate.

\subsection{School matching}
\label{sec:schoolmatching}
Finally, consider school matching. In many urban environments, students are assigned to public schools through the deferred acceptance algorithm \cite{abdulkadiroglu_new_2005}. The algorithm inputs applicant preferences (via ranked lists of schools) and school priorities (with factors such as geography, academic performance, diversity, and lottery numbers). The promise is twofold: (a) these systems provide the opportunity to access desired schools, even if they are not in the student's neighborhood; (b) when slots in high-value schools are scarce, they are allocated not solely due to geography but also accounting for student preferences, academic performance, and random chance---thus, these allocation systems can be more effective and equitable than those that simply reflect geographic segregation. 

In practice, applying effectively can be time consuming for families: in NYC, there are over 800 high school programs to choose from, each with varying locations, classes and sports teams offered, and school quality metrics. Families who can afford it often pay for admissions consultant services. A long line of research has empirically shown that information access, awareness, and the time-consuming process -- not just preferences -- affects application behavior, both in NYC high school admissions \cite{corradini2023information,idoux2024overcoming} and elsewhere \cite{larroucau2024college,tomkins2023showing,arteaga2022smart,ajayi2020school}.

The ``administrative burdens'' \cite{herd2019administrative} of applying lead to participation heterogeneity and outcome inequity. For example, \citet{cohodes2022informational} documents the large fraction of students who apply to non-competitive, ``nonoptimal'' schools first in their rankings. In \citet{peng2025deviations}, we show that such behavior leads to substantial ``undermatching'': students not matching to as high-quality programs as they could have (that are no further geographically than their actual match), because they did not apply. In particular, this gap between where students matched and where they could have matched is almost twice as large for the most competitive Black and Hispanic students as it is for the corresponding Asian and white ones, when quality is measured by program performance, value add, selectivity, school graduation rate, or college enrollment rate. We then show that simple application behaviors explain a large portion of this undermatching. 
Including with surveys, \citet{idoux2024overcoming} show that differential awareness of schools rated as high-quality and racial homophily preferences explain such gaps, as opposed to preferences over other  characteristics like quality.

Substantial work has further gone into developing and evaluating \textit{informational interventions} to close the participation gaps \cite{Corcoran18,arteaga2022smart,cohodes2022informational,corradini2023information,larroucau2024college}---for example, by providing students lists of high-quality programs close to their neighborhood. These interventions have changed behavior, when used. 

Despite this focus, much work remains to be done.  Informational gaps and heterogeneous participation persist, as documented by recent studies  \cite{idoux2024overcoming,peng2025deviations}. One constant challenge, highlighted by  \citet{cohodes2022informational}, is that informational interventions only work to the extent that they are \textit{used}, i.e., they put the burden on participants---just as targetted advertising for participatory budgeting ultimately requires people to respond to the ads. Thus, it remains open how to deploy (a) interventions with high takeup rates and (b) new mechanisms robust to heterogeneous participation.

\subsection{Common Themes and Implications}
The above examples all follow a similar pattern: a mechanism allocates scarce public resources or makes joint decisions; a core mechanism component is to input preferences or information from participants; when the mechanism is deployed, participation is heterogeneous, despite it being monetarily ``free.'' Such heterogeneity both makes the mechanism less effective and potentially skews allocation and decisionmaking against those already disadvantaged. While substantial work has been done to measure and reduce these disparities, they persist.

Related concerns potentially apply in other settings in which preferences are elicited from participants who may have heterogeneous capabilities: in refugee matching, refugees may be asked for preferences over host countries \cite{jones2017international}; in food bank allocation, large food banks (but not small ones) have dedicated staff to interface with the mechanism \cite{prendergast2017food}; in kidney exchange, preference elicitation from doctors regarding compatible kidneys is a practical challenge \cite{ashlagi2021kidney}. 

What should we do, given this fact pattern? In any given setting, the options are to (a) defend the status quo, by establishing that the mechanism is nevertheless effective, or at least preferable over any feasible counterfactual mechanism; (b) aim to reduce participation heterogeneity, as behavior is far from fixed; (c) reform the mechanism, so that it is robust; and/or (d) replace it entirely, likely to one that minimally uses the people's preferences and information. How should we choose which option(s) to pursue? Different applications have and should take different paths, and the paths are complementary.\footnote{There are key differences between the applications. Voting leads to a collective decision, and individuals who do not participate nevertheless benefit if they agree with those who do. In stable matching, allocations are individual and  more arguably `zero-sum.' Crowdsourcing lies in-between, as allocations (e.g., pothole fixes) are geographically localized, but everyone may benefit from the information shared by participants. The exposition has ignored these differences, as they are not crucial to my core thesis. However, they may be relevant in considering paths forward.} In the face of heterogeneous participation, we defend standard democracy \textit{and} invest substantially in voter turnout efforts; this is also the path taken so far for participatory budgeting and deliberative democracy. In other cases, we've seen either reform or a retreat from participatory mechanisms. 

I posit that reform should aim for the following ``best of both worlds'' north star: \textit{use preferences and information from those who choose to participate, but provide a ``sufficient'' quality of service to those who do not.}  In other words, we should attempt to retain the benefits of public participation while mitigating the resulting resource allocation skews. With this goal, we would still have ``power users'' who  benefit from their invested time; however, non-participation would lead to a reasonable, default allocation. Of course, ``sufficient'' and ``reasonable'' are subjective, and themselves policy choices; when allocating scarce resources, these defaults may come at some cost to the ``power users.'' Policymakers and the public are in the best position to choose the context-dependent operating point on the inevitable tradeoff between information aggregation and allocation skews. My position is that this choice should be explicit, as opposed to the too-common status quo of maximally supporting aggregation at the cost of allocation skews. 

\section{Approaches for practice}

In the remainder of this article, I highlight potential directions for practice and research, in service of this goal. These paths are informed by the above literature and \citet{herd2019administrative} in particular. They lay out three reforms to respond to the ``Medicare Maze,'' in which the elderly must annually learn about complex options to choose a health care plan, leading to worse health outcomes and increased costs: (1) reduce choice by simplifying options; (2) expand outreach and human assistance in navigating the choices; (3) use administrative data and information technology to provide personalized defaults or recommendations. These options have their analogues for participatory mechanisms.

\subsection{Reduce participation heterogeneity} The simplest response to participation heterogeneity is to try to reduce it. In participatory budgeting, this is done via targeted advertising; in deliberative democracy, this is done more directly by modifying selection probabilities.\footnote{This is only possible because deliberative democracy purposely is designed to select a subset of the people, with the goal of making that subset representative.} This approach is also a key tool to reduce disparities in the takeup of other entitlements, like SNAP benefits in the United States; \citet{koenecke2023popular} show public support for targeted advertising that improves allocation equity. However, as continued disparities prove, turnout efforts are not a panacea in the presence of structural barriers to participation, such as those discussed by \citet{mullainathan2013scarcity}. Approaches that more directly tackle structural barriers, such as those that provide childcare and video conferencing technology for deliberative democracy, may be necessary.

Another, more systematic approach to reducing participation \textit{heterogeneity} is to use preferences \textit{within} areas with relatively homogeneous participation. For example, NYC runs participatory budgeting separately for each city council district, with a set budget per district; \textit{if} districts are drawn such that participation is similar within each district, then heterogeneity across districts would not skew allocations.\footnote{It is not clear that NYC's districts meet this criteria. My district spans relatively wealthy areas in the Upper West Side, to Columbia University,  to lower-income areas in West Harlem. However, granular turnout data is unavailable and  winning projects did not geographically concentrate.} 
Analogously, in the tree planting context, the following approach could incorporate geographic balance, need as determined by the agency, and resident requests: make neighborhood-level scheduling and quantity decisions according to agency expertise; then, within each neighborhood, allow requests to inform precise planting locations, alongside expertise. Appropriately designed, such an approach could be ``best of both worlds'' and combine elicited preferences with expert decisionmaking.

\subsection{Provide personalized defaults or recommendations} Turnout campaigns may be effective when participation is (meaningfully approximated as) binary: in voting mechanisms and resident crowdsourcing, the most important outcomes are how regularly someone votes or submits requests. In school matching, on the other hand, \textit{whether} people submit ranked lists is not the only concern, as doing so is required to enroll a child in public school. Rather, submitting \textit{informed} ranked lists is a challenge, as it requires awareness of program quality and admissions probabilities; only some may have access to expensive consulting services or advice from social networks to help them navigate these decisions. In such systems, practitioners, alongside researchers, may be able to provide recommendations or even default options to users. Then, applicants can -- just as in the status quo -- provide preferences if they are dissatisfied with the recommendations or defaults; others can choose to follow the recommendations. Of course, as with targetted advertising, one challenge with recommendations is takeup \cite{cohodes2022informational}, and so stronger user interfaces or  ``nudges'' are important.

In many cases, as in school matching, there already \textit{is} a default option, e.g., a manual administrative placement if an applicant does not match with any school. One approach is for these defaults to be more systemically planned, to provide better allocations to those who do (can) not participate meaningfully. Recommendations and defaults are also related to -- and `lighter-touch' than -- another approach developed in school matching: limiting options, potentially in a data-driven manner: \citet{shi2015guiding} develops short choice menus for each family in Boston, citing ``too many options'' as contributing to long commute times, unpredictability, loss of neighborhood cohesion, and a research burden on families; \citet{allman2022designing} develop small zones in San Francisco, in support of school diversity.

The use of personalized defaults and recommendations, powered by modern machine learning methods, may also be effective in other contexts. \citet{ashlagi2021kidney} advocate for a related approach in the context of preference elicitation difficulties for kidney exchanges: ``it may be useful to develop machine learning models to predict positive crossmatches and ... to understand the trade-offs involved with
waiting (while on dialysis) for a better match.''

\subsection{Actively acquire information or post-process inputs}
A third approach is for the central system to actively invest in information acquisition to counter participation biases. When resident crowdsourcing informs resource allocation, for example, agencies can invest more active inspection resources or install sensors (e.g., flood sensors \cite{franchi2025bayesian}) in neighborhoods with lower reporting rates. 
Alternatively, given the public's heterogeneous inputs, the system can make decisions that are nevertheless balanced. As an example of such post-processing, consider our work with the New York Public Library on the holds system, which allows patrons to request books from any system branch to be sent to their local neighborhood branch; we first found that heterogeneous usage of the holds system (even conditional on overall library usage) led to a large net outflow of books from lower-income neighborhoods to higher-income ones \cite{liu2024identifying}. We then designed a routing prioritization scheme between branches to mitigate such disparities \cite{liu2025optimizing}, so that all holds requests could be fulfilled without disproportionately depleting branches in lower holds-use neighborhoods.

However, these approaches are not always feasible. In deferred acceptance, where applicant preferences are directly used, it is unclear where active information acquisition can be incorporated or how matches can be post-processed. In democratic systems such as participatory budgeting, weighting votes may conflict with other design principles, such as `one-person-one-vote' (as discussed by \cite{gelauff2022opinion}). Such approaches may be feasible when constructing error bars or using vote outcomes to advise final decision-makers; however, the question of ``representativeness'' (and \textit{of whom}) remains, especially when participation correlates with unobserved features \cite{chasalow2021representativeness}. 

These solutions are analogous to those proposed in algorithmic fairness, to counter disparities in prediction accuracy that are caused by heterogeneous unobserved confounding or missing data. There, data may be actively acquired or post-processed while using demographics as features \cite{chen2018my,noriega2019active,caton2020fairness,cai2020fair,garg2021standardized,liu2021test,movva2023coarse,zink2024race,balachandardomain,dong2025addressing,chiang2025bayesian}. There as well, post-processing may be infeasible, due to legal constraints or a general preference for ``group-unaware'' approaches (e.g., the recent affirmative action ban in college admissions in the United States, which also affect algorithms in the admissions process \cite{lee2024ending}). More generally, I believe that the goal of countering heterogeneous participation may further connect market design to algorithmic fairness, cf. \citet{finocchiaro2021bridging}.

\vspace{.5em}
\noindent All three approaches use central resources to counter  heterogeneous participation and pursue ``best of both words'': use elicited preferences, but mitigate allocation skews. Next, I discuss how researchers can contribute to the vision.

\section{Research directions}
Researchers have an important role to play in collaborating with practitioners on designing, deploying, and evaluating the above approaches. Researchers---including those who do not collaborate with practitioners---can also contribute in other ways. Below, I overview three approaches for a diverse range of skillsets: (a) empirically quantifying heterogeneous participation; (b) providing theoretical insight on \tradeoff\ and designing mechanisms to navigate it; (c) more directly considering human-computer-market interactions and interface design.

\subsection{Empirically quantify heterogeneous participation} Academics -- through open data, information requests, or practitioner collaboration -- can help quantify participation heterogeneity. Methodologically, the challenge is that quantifying participation heterogeneity often requires disambiguating it from other, less concerning, explanations. In resident crowdsourcing, we must show that heterogeneous \textit{conditions} cannot explain the discrepancy -- that it is not the case that some neighborhoods report less {because} they encounter fewer incidents worth reporting \cite{liu2024quantifying,agostini2024bayesian,balachandar2024using,franchi2025bayesian}. In school matching, we must show that heterogeneous preferences -- e.g., due to outside options or true heterogeneity in idiosyncratic preferences for certain schools or school characteristics -- do not fully explain behavior, and that instead heterogeneous information plays an important role \cite{larroucau2024college,idoux2024overcoming,corradini2023information}. This challenge often requires new statistical methods, analyzing natural experiments, or careful collection of ``ground truth'' data, such as through surveys and randomized controlled trials. 

Such quantification helps provide an empirical underpinning with which interventions can be justified and well-engineered. For example, quantifying missing incident reports by neighborhood helps in the allocation of inspection and sensor resources in resident crowdsourcing, and quantifying heterogeneous awareness and behavior informs the design of personalized recommendations in school matching.

Finally, I note that empirically quantifying heterogeneous participation is related to two empirical lines of work: (1) preference estimation under strategic behavior \cite{agarwal2018demand,Calsamigliademand}, where the goal is to estimate preferences in non-strategyproof mechanisms, when (some) agents may be strategic; (2) empirical behavioral economics, that seeks to quantify how human behavior deviates from ``optimal,'' including in strategyproof mechanisms. Here, my focus is on quantifying \textit{heterogeneous} behavior and its effects on downstream resource allocation, especially when there is no formal cost or strategic incentive.

\subsection{Theoretically model allocation under heterogeneous participation and design mechanisms to explicitly navigate the participation-allocation tradeoff}

Theoretical modeling of heterogeneous participation is a rich area for further study, to complement empirical measurement. Models can (a) elucidate  welfare outcomes under heterogeneous participation; and (b) help design better mechanisms.

In the context of school matching, \citet{kloosterman2020school} analyze a setting in which some students are more informed than others about high quality options; under the model, such students may be \textit{worse off} under deferred acceptance than without school choice; they then show that priorities may be designed in a way to avoid this outcome. \citet{pathak2008leveling} analyze matching settings in which some students are ``sophisticated'' (strategic), while others are sincere despite strategic incentives; while ``sincere students lose priority to sophisticated students'' under the non-strategy-proof Boston mechanism, ``any sophisticated student weakly prefers her assignment under the Pareto-dominant Nash equilibrium of the Boston mechanism to her assignment under the recently adopted student-optimal stable mechanism.'' It is essential to develop such models for other settings, as well as experiment with and deploy mechanisms with properties similar to the ones developed by \citet{kloosterman2020school}. More generally, some mechanisms may be more effective at supporting diverse participation. 

One setting where such conceptual insights helped was the design of Feeding America's market mechanism to allocate food to food banks. As detailed by \citet{prendergast2017food}, an essential consideration was to protect smaller food banks from heterogeneous participation, as they have ``fewer resources and manpower ... relative to their larger counterparts, where there are often dozens of workers or volunteers.'' The chosen mechanism avoided a continuous auction (which would benefit those with dedicated staff members) and allowed fractional bidding and storing of credits. It further effectively enabled a default option, giving food banks ``the option to delegate bidding to an employee of Feeding America, where a food bank could simply outline in broad terms its needs to that person'' \cite{prendergast2017food}.

A related question suitable for modeling insight is: under what contexts is the \tradeoff\  big, and when should we potentially abandon a mechanism? This question has recently been explored in the context of individual-level {prediction} to target resources: \citet{shirali2024allocation} argue that ``prediction-based allocations outperform baseline methods using aggregate unit-level statistics only when between-unit inequality is low and the intervention budget is high,'' i.e., that the cost of individualized predictions may not be worth it; \citet{perdomo2023difficult} empirically illustrate such ideas in the context of targeting interventions for students at risk of dropping out of school. \citet{wang2024against} argue against the legitimacy of decisionmaking that uses predictions of the future about individuals, due to reoccurring challenges regarding accuracy, disparate performance, and related concerns. Analogously, it may emerge in a model that eliciting preferences is only worth it when heterogeneity from preferences is larger than that from participation.

\subsection{More directly consider human-computer-market interaction} Finally, the EconCS community should increase collaborations with human-computer interaction (HCI) researchers, to build {interfaces} that more effectively allow equal participation. \citet{schank2021power} and \citet{pahlka2023recoding} both pinpoint {bad} interface design as worsening government service. I posit that (1)  good interfaces may be more effective than good theoretical properties in improving participation and systems, and (2) qualitative studies are important to understand participation. Here, I briefly overview my work and collaborations with HCI researchers. 

In \citet{bartle2025faster}, we build and deploy an SMS-based system to help place patients being discharged from hospitals into nursing care homes. In our context in Hawai`i, care homes are often run by retired nurses \textit{out of their own homes}, with only one or two patients; whenever a patient needs to be placed, a full-time team of hospital social workers calls the approximately thousand nursing homes to see if they have capacity and can care for the given patient's needs. This preference information is not centrally available because integration into a healthcare management system like Epic does not work for this rural, single-operator population. As we show, simply collecting capacity information -- through SMS -- and showing the data to hospital social workers trying to place patients into homes improves the process; my conjecture is that this data improvement -- enabled by effective interface design for care home operators -- is far greater than improved call recommendations, such as through a matching optimization, would yield. However, one challenge is that many homes do not share their preferences over patient characteristics with the system. In followup work, we ran a randomized controlled trial and interviewed care home operators to understand this participation gap \cite{bartle2025shopping}. The experiment revealed that nudges can (somewhat) increase the number of homes who share their preferences, and the interviews uncovered complex cultural phenomena as well as economic considerations that shape dynamic preferences. The mixed methods approach and collaboration across fields was essential in understanding participation and how interventions may increase it. 

Substantial work has also shown that interface design can affect behavior in other systems studied by the community. In participatory budgeting, substantial work compares the behavioral and learning implications of the elicitation mechanism (i.e., whether they voters asked to rank or approve projects, or to modify a proposed budget or give a full budget) \cite{gelauff2018comparing,garg2019iterative,garg2019your,goel2019knapsack,gelauff2024rank}---with the hypothesis that some mechanisms may be easier for voters to understand. In ratings systems, simple modifications such as the question that is asked can substantially affect behavior, by aligning different people on what ``five stars'' actually means \cite{garg2019designing,garg2021designing}. Similarly, I conjecture that interface design could reduce participation gaps in other systems. While simplifying interfaces is likely to be generally useful, open questions remain on how to best present information, including recommendations. Future work should experimentally evaluate interfaces and qualitatively interview participants regarding how they perceive a given interface and system design.

\section{Conclusion}

Economic and computational researchers have important roles to play in designing and analyzing societal systems \cite{roth2002economist,abebe2020roles}. Our community should be proud of our  impact in influencing the deployment of so many real-world systems. Undoubtedly, many of these systems improve upon those that they replaced. However, just as we theoretically design mechanisms to be strategyproof, so that people \textit{can safely} share their true preferences, we should focus on whether people \textit{do} participate on equal footing, or can do so in the presence of heterogeneous time costs. We should further engineer our systems -- theoretically, algorithmically, and through interface design -- so that they do not inadvertently allocate scarce resources according to participation ability. In this article, I overviewed research, including much of my own, in pursuit of this goal. I believe that ``best of both worlds'' systems, that incorporate preferences without allowing heterogeneous participation to skew distributional outcomes, are possible and necessary.

\section*{Acknowledgments}
I thank Rediet Abebe, Bailey Flanigan, Lodewijk Gelauff, Paul G{\"o}lz, Allison Koenecke, Karen Levy, Irene Lo, and members of Cornell's AI, Policy, and Practice Initiative for invaluable discussion and feedback, as well as my advisors, students, and collaborators. NG is supported by NSF CAREER IIS-2339427, and Cornell Tech Urban Tech Hub, Meta, and Amazon research awards.

\bibliographystyle{plainnat}
\bibliography{ng_bib}

\end{document}